\documentclass{ws-ijmpe}
\usepackage[super,compress]{cite}

\begin{document}

\markboth{M. Thoennessen}{2016 Update of the Discoveries of Isotopes}

\catchline{}{}{}{}{}

\title{2016 UPDATE OF THE DISCOVERIES OF NUCLIDES}

\author{\footnotesize M. THOENNESSEN}

\address{National Superconducting Cyclotron Laboratory and \\
Department of Physics \& Astronomy \\
Michigan State University\\
East Lansing, Michigan 48824, USA\\
thoennessen@nscl.msu.edu}

\maketitle

\begin{history}
\received{Day Month Year}
\revised{Day Month Year}
\end{history}

\begin{abstract}
The 2016 update of the discovery of nuclide project is presented. Only twelve new nuclides were observed for the first time in 2016. A large number of isotopes is still only published in conference proceedings or internal reports. No changes to earlier assignments were made.
\end{abstract}


\ccode{PACS numbers: 21.10.-k, 29.87.+g}


\section{Introduction}

This is the fourth update of the isotope discovery project which was originally published in a series of papers in Atomic Data and Nuclear Data Tables from 2009 through 2013 (see for example the first \cite{2009Gin01} and last \cite{2013Fry01} papers). Two summary papers were published in 2012 and 2013 in Nuclear Physics News \cite{2012Tho03} and Reports on Progress in Physics \cite{2013Tho02}, respectively, followed by annual updates in 2014 \cite{2014Tho01}, 2015 \cite{2015Tho01} and 2016 \cite{2016Tho02}. The 2016 update included an overall reevaluation to apply the criteria uniformly for all elements.  A description of the discoveries from an historical perspective was published last year in the book ``The Discovery of Isotopes -- A complete Compilation''  \cite{2016Tho01}.

\section{New discoveries in 2016}
\label{New2016}

In 2016, the discoveries of twelve new nuclides were reported in refereed journals. This includes the first observation of an unbound resonance in the tetra-neutron system. In addition, eight nuclides along the proton-dripline and three transuranium nuclides were discovered. Table \ref{2016Isotopes} lists details of the discovery including the production method. With the exception of $^{178}$Pb all isotopes were identified at RIKEN in Japan.

\begin{table}[pt]
\tbl{New nuclides reported in 2016. The nuclides are listed with the first author, submission date, and reference of the publication, the laboratory where the experiment was performed, and the production method (PF = projectile fragmentation, FE = fusion evaporation, SB = secondary beams). \label{2016Isotopes}}
{\begin{tabular}{@{}llrclc@{}} \toprule 
Nuclide(s) & First Author & Subm. Date & Ref. & Laboratory & Type \\ \colrule
$^{4}$n  & K. Kisamori & 7/30/2015 &\refcite{2016Kis01}& RIKEN & SB \\
$^{96}$In, $^{94}$Cd, $^{92}$Ag, $^{90}$Pd & I. Celikovic & 1/19/2016 & \refcite{2016Cel01} & RIKEN & PF  \\ 
$^{63}$Se, $^{67}$Kr, $^{68}$Kr & B. Blank & 4/11/2016 & \refcite{2016Bla01} &  RIKEN & PF \\
$^{178}$Pb &  H. Badran & 6/9/2016 & \refcite{2016Bad01} &  Jyv\"askyl\"a & FE \\
$^{230}$Am, $^{234}$Cm, $^{234}$Bk & D. Kaji & 9/29/2016 & \refcite{2016Kaj01} &  RIKEN & FE \\
\botrule
\end{tabular}}
\end{table}

Evidence for the observation of the four-neutron resonance was reported by Kisamori et al. in the paper ``Candidate resonant tetraneutron state populated by the $^4$He($^8$He,$^8$Be) reaction'' \cite{2016Kis01}. The state was detected in the missing-mass spectrum following the double-charge-exchange reaction with a 186 MeV/u radioactive $^8$He beam from the Radioactive Ion Beam Factory (RIBF) of the RIKEN Nishina Center. The tentative claim for a bound tetra-neutron in 2002 \cite{2002Mar01} was not confirmed in subsequent experiments \cite{2005Ale01}. In addition, parts of the analysis were questioned \cite{2004She01} and the existence of a bound tetraneutron seems theoretically very unlikely \cite{2003Pie01}.

$^{96}$In, $^{94}$Cd, $^{92}$Ag, and $^{90}$Pd were discovered by \v{C}elikov\'ic et al. and reported in ``New Isotopes and Proton Emitters--Crossing the Drip Line in the Vicinity of $^{100}$Sn'' \cite{2016Cel01}. A 345 MeV/A primary $^{124}$Xe beam from RIBF was fragmented on thick $^9$Be targets and separated with the BigRIPS projectile fragment separator. The isotopes were identified using the $\Delta E-TOF-B\rho$ method. ``Four new isotopes, namely, $^{96}$In, $^{94}$Cd, $^{92}$Ag, and $^{90}$Pd, have been clearly identified with 2, 3, 8, and 2 events, correspondingly.''

Blank et al. reported the observation of $^{63}$Se, $^{67}$Kr, and $^{68}$Kr in ``New neutron-deficient isotopes from $^{78}$Kr fragmentation'' \cite{2016Bla01}. A 345 MeV/A primary $^{78}$Kr beam from RIBF was fragmented on thick $^9$Be targets and separated with the BigRIPS projectile fragment separator. The isotopes were identified using the $\Delta E-TOF-B\rho$ method. ``Clean identification spectra could be produced and $^{63}$Se, $^{67}$Kr, and $^{68}$Kr were identified for the first time.''

The first identification of $^{178}$Pb was published by Badran {\it et al.} in the paper entitled ``Confirmation of the new isotope $^{178}$Pb'' \cite{2016Bad01}. A self-supporting $^{104}$Pd target was bombarded with a 358~MeV $^{78}$Kr beam from the K-130 cyclotron at Jyv\"askyl\"a. The recoils from the fusion evaporation reaction $^{104}$Pd($^{78}$Kr,4n)$^{178}$Pb were separated with the gas-filled separator RITU and identified in the GREAT spectrometer. ``The half-life of the ground state of $^{178}$Pb was determined to be $T_{1/2} = 0.21^{+0.21}_{-0.08}$~ms using the maximum-likelihood method.'' The word confirmation in the title of the paper referred to a previous observation of $^{178}$Pt which, however, was only published as a conference proceeding \cite{2003Bat01}. 

$^{230}$Am, $^{234}$Cm, and $^{234}$Bk were discovered by Kaji et al. as reported in the paper ``Decay Properties of New Isotopes $^{234}$Bk and $^{230}$Am, and Even-Even Nuclides $^{234}$Cm and $^{230}$Pu'' \cite{2016Kaj01}. A 189.5 MeV $^{40}$Ar beam accelerated by the RIKEN heavy-ion  linear accelerator RILAC bombarded a gold target to form $^{234}$Bk in the reaction $^{197}$Au($^{40}$Ar,3n). Reaction products were separated with the gas-filled ion separator GARIS and transported to the rotating wheel system MANON where correlated $\alpha$-particles and fission fragments were measured. ``Alpha-decay energies of eleven $^{234}$Bk were found at 7.62$-$7.96~MeV, and six fission events that correlated with the $\alpha$-decay of $^{234}$Bk were observed. The half-lives of $^{234}$Bk and $^{230}$Am were determined to be 19$^{+6}_{-4}$~s and 3$^{+22}_{-9}$~s, respectively. The $^{234}$Cm followed by the $\beta$-decay of $^{234}$Bk was also identified.'' The observation of $^{234}$Cm was not considered a new discovery referring to an internal report \cite{2002Cag01} and a conference proceeding \cite{2002Cag02}.

In addition to these new discoveries in 2016, the observation of $^{215}$U was reported in 2015 but had not been included in the  update of the compilation last year. It was reported by Yang et al. in the paper ``Alpha decay of the new isotope $^{215}$U'' \cite{2015Yan01}. $^{215}$U was formed in the fusion evaporation reaction $^{180}$W($^{40}$Ar,5n)$^{215}$U with an 205.5 MeV $^{40}$Ar beam delivered from the Sector-Focusing Cyclotron of the Heavy Ion Research Facility in Lanzhou, China. The gas-filled recoil separator for Heavy Atoms and Nuclear Structure (SHANS) was used to separate evaporation residues which were implanted in a position-sensitive silicon strip detector (PSSD). $^{215}$U was identified by detecting correlated $\alpha$ particles in the PSSD or in a box of eight silicon detectors surrounding the PSSD in the backward direction. ``The $\alpha$-particle energy and half-life of $^{215}$U were determined to be 8.428(30)MeV and 0.73$^{+1.33}_{-0.29}$~ms, respectively.''

\section{Status at the end of 2016}

The twelve new discoveries in 2016 plus the inclusion of the 2015 discovery of $^{215}$U  increased the total number of observed isotopes to 3224. They were reported by 905 different first authors in 1537 papers and a total of 3667 different coauthors.  Further statistics can be found on the discovery project website \cite{2011Tho03}.

\begin{figure}[pt] 
\centerline{\psfig{file=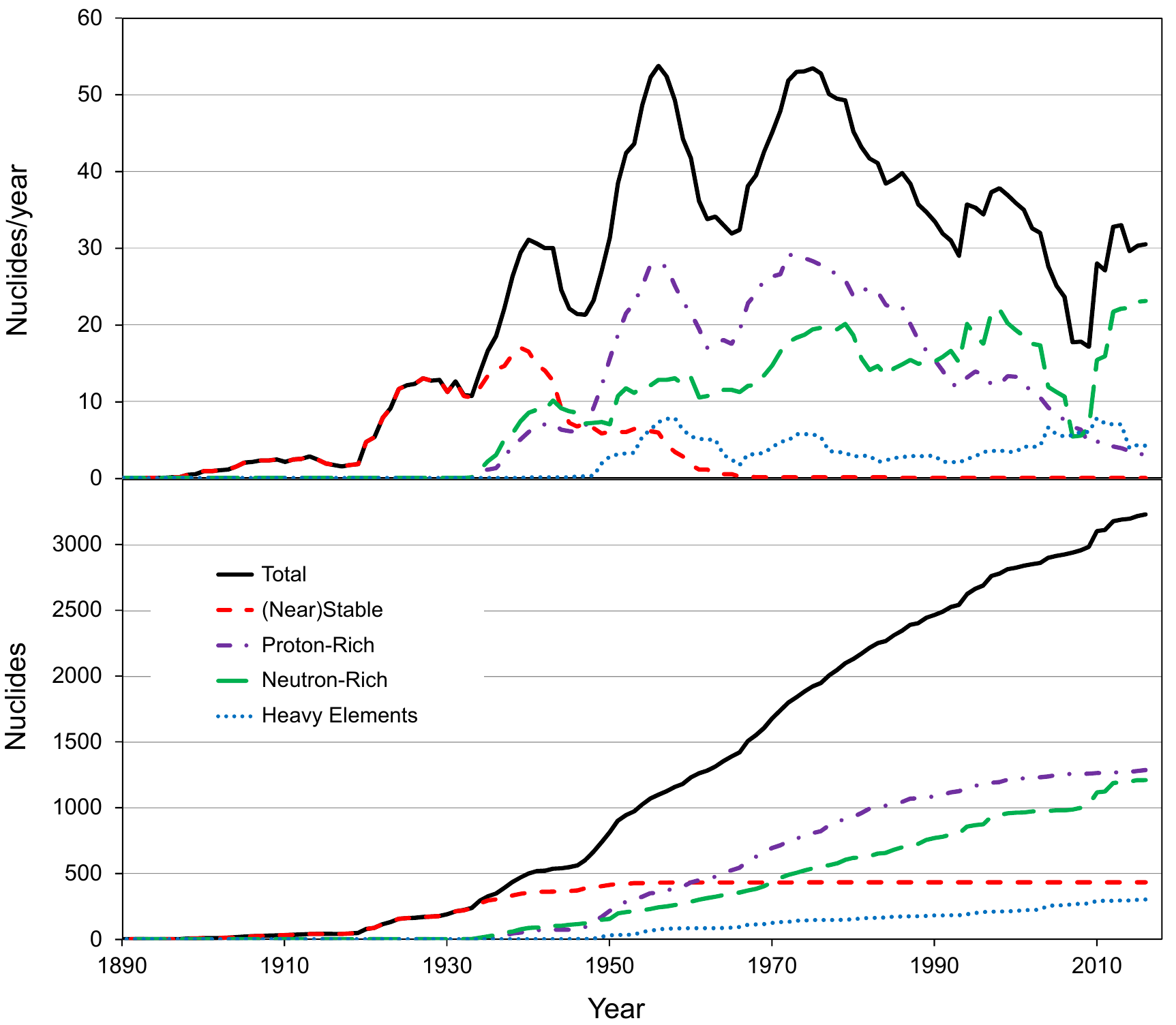,width=12.4cm}}
\caption{Discovery of nuclides as a function of year. The top figure shows the 10-year running average of the number of nuclides discovered per year while the bottom figure shows the cumulative number.  The total number of nuclides shown by the black, solid lines are plotted separately for near-stable (red, short-dashed lines), neutron-deficient (purple, dot-dashed lines), neutron-rich (green, long-dashed lines) and transuranium (blue, dotted lines) nuclides. This figure was adapted from Ref. 5 to include the data from 2016. \label{f:timeline} }
\vspace*{-0.1cm}
\end{figure}

Figure \ref{f:timeline} shows the current status of the evolution of the nuclide discoveries for four broad areas of the nuclear chart, (near)stable, proton-rich, neutron-rich, and the region of the heavy elements. The figure was adapted  from the 2014 review\cite{2014Tho01} and was extended to include all isotopes discovered until the end of 2016. The top part of the figure shows the ten-year average of the number of nuclides discovered per year while the bottom panel shows the integral number of nuclides discovered. 

Although the ten-year average rate for neutron-rich isotopes reached an all-time high of 23.1, overall there are still more known proton-rich than neutron-rich nuclides. The total number of isotopes discovered by projectile fragmentation or projectile fission crossed 600. The method is used to produce neutron- as well as proton-rich nuclides. It corresponds to the second largest production mechanism only behind fusion-evaporation reactions which account for over 750 isotope discoveries.

\section{Discoveries not yet published in refereed journals}

Eight  isotopes ($^{96}$In, $^{94}$Cd, $^{92}$Ag, $^{90}$Pd, $^{178}$Pb, $^{230}$Am, $^{234}$Cm, and $^{234}$Bk) included in the list of isotopes only reported in proceedings or internal reports in last year's update \cite{2016Tho02} have been published in refereed publications this year. While $^{215}$U was still included in this list it was actually published in a refereed journal in 2015 \cite{2015Yan01}. 

Table \ref{reports} lists the isotopes which so far still have only been presented in conference proceedings or internal reports. This includes almost 70 isotopes produced by projectile fragmentation and projectile fission at RIBF which should be published in refereed journals in the near future. The table includes two new entries: (i) $^{280}$Ds which was presented at  the second Conference on Advances in Radioactive Isotope Science (ARIS) in 2014 \cite{2014Mor01} and has been discussed in the 2014 RIKEN accelerator report \cite{2015Mor01}, and (ii) a Ph.D. thesis from 2005, reporting an additional measurement of $^{255}$Db \cite{2005Lep01}.

Seven isotopes ($^{116}$Nb, $^{138}$In, $^{143}$Sb, $^{145}$Te, $^{147}$I, $^{149}$Xe, and $^{154}$Ba) presented at the 4$^{th}$ Joint Meeting of the APS Division of Nuclear Physics and the Physical Society of Japan in 2014 \cite{2014Shi01} have not been included in a proceeding or annual report so far.

\begin{table}[t]
\tbl{Nuclides only reported in proceedings or internal reports until the end of 2016. The nuclide, first author, reference and year of proceeding or report are listed. \label{reports}}
{\begin{tabular}{@{}llrr@{}} \toprule
\parbox[t]{6.8cm}{\raggedright Nuclide(s) } & \parbox[t]{2.3cm}{\raggedright First Author} & Ref. & Year \\ \colrule

$^{20}$B		&	 F. M. Marqu\'es 	&	\refcite{2015Mar01}	&	2015	 \\ 
$^{21}$C		&	 S. Leblond 	&	\refcite{2015Leb01}	&	2015	 \\ 
$^{86}$Zn, $^{88}$Ga, $^{89}$Ga, $^{91}$Ge, $^{93}$As, $^{94}$As, $^{96}$Se, $^{97}$Se	&	 Y. Shimizu 	&	\refcite{2015Shi01}	&	2015	 \\ 
$^{99}$Br, $^{100}$Br & & & \\
$^{81}$Mo, $^{82}$Mo, $^{85}$Ru, $^{86}$Ru	&	 H. Suzuki	&	\refcite{2013Suz01}	&	2013	 \\ 
$^{98}$Sn, $^{104}$Te	&	  I. Celikovic 	&	\refcite{2013Cel01}	&	2013	 \\ 
$^{122}$Tc, $^{125}$Ru, $^{130}$Pd, $^{131}$Pd, $^{140}$Sn, $^{142}$Sb, $^{146}$I	&	 Y. Shimizu 	&	\refcite{2014Shi02}	&	2014	 \\ 
$^{153}$Ba, $^{154}$La, $^{155}$La, $^{156}$Ce, $^{157}$Ce, $^{156}$Pr$^a$, $^{157}$Pr, 	&	 D. Kameda 	&	\refcite{2013Kam01}	&	2013	 \\
$^{158}$Pr, $^{159}$Pr, $^{160}$Pr, $^{162}$Nd, $^{164}$Pm, $^{166}$Sm & & & \\
$^{155}$Ba, $^{157}$La, $^{159}$Ce, $^{161}$Pr, $^{164}$Nd, $^{166}$Pm, $^{168}$Sm,  	&	 N. Fukuda 	&	\refcite{2015Fuk01}	&	2015	 \\
$^{170}$Eu, $^{172}$Gd, $^{173}$Gd, $^{175}$Tb, $^{177}$Dy, $^{178}$Ho, $^{179}$Ho,& & & \\
 $^{180}$Er, $^{181}$Er, $^{182}$Tm, $^{183}$Tm & & & \\
	$^{156}$La, $^{158}$Ce, $^{163}$Nd, $^{165}$Pm, $^{167}$Sm, $^{169}$Eu, $^{171}$Gd,	&	 D. Kameda 	&	\refcite{2014Kam02}	&	2014	 \\
$^{173}$Tb, $^{174}$Tb, $^{175}$Dy, $^{176}$Dy, $^{177}$Ho, $^{179}$Er & & & \\
$^{126}$Nd, $^{136}$Gd, $^{138}$Tb, $^{143}$Ho$^b$, $^{150}$Yb, $^{153}$Hf	&	 G. A. Souliotis 	&	\refcite{2000Sou01}	&	2000	 \\
	$^{143}$Er, $^{144}$Tm	&	 R. Grzywacz 	&	\refcite{2005Grz01}	&	2005	 \\
	& K. Rykaczewski & \refcite{2005Ryk01} & 2005 \\
	& C. R. Bingham & \refcite{2005Bin01} & 2005 \\
$^{230}$At, $^{232}$Rn	&	 J. Benlliure 	&	\refcite{2010Ben02}	&	2010	 \\
$^{235}$Cm	&	 J. Khuyagbaatar 	&	\refcite{2007Khu01}	&	2007	 \\
$^{252}$Bk, $^{253}$Bk	&	 S. A. Kreek 	&	\refcite{1992Kre01}	&	1992	 \\
$^{262}$No 	&	 R. W. Lougheed 	&	\refcite{1988Lou01},\refcite{1989Lou01}	&	1988/89	 \\
	&	 E. K. Hulet 	&	\refcite{1989Hul01}	&	1989	 \\
$^{261}$Lr, $^{262}$Lr	&	 R. W. Lougheed 	&	\refcite{1987Lou01}	&	1987	 \\
	&	 E. K. Hulet 	&	\refcite{1989Hul01}	&	1989	 \\
	&	 R. A. Henderson 	&	\refcite{1991Hen01}	&	1991	 \\
$^{255}$Db	&	 G. N. Flerov	&	\refcite{1976Fle01}	&	1976	 \\
		& A.-P. Lepp\"anen	& \refcite{2005Lep01} & 2005 \\
$^{280}$Ds	&	 K. Morita	&	\refcite{2015Mor01}	&	2014	 \\
\botrule
\vspace*{-0.2cm} & & & \\
$^a$ also published in ref. \refcite{1996Cza01} and \refcite{1997Ber02} \\
$^b$ also published in ref. \refcite{2003Sew02} \\
\end{tabular}}
\end{table}

\section{Summary}

Following the revisions performed in preparations for the book `` The Discovery of Isotopes ''  \cite{2016Tho01} presented in last year's update \cite{2016Tho02} no major changes to the discovery assignments were made. As the example of $^{215}$U demonstrates, sometimes discoveries are overlooked. Thus continued input, feedback, and comments from researchers are encouraged to ensure that the compilation is always complete and up-to-date.

\section*{Acknowledgements}

Support of the National Science Foundation under grant No. PHY15-65546 is gratefully acknowledged.


\bibliographystyle{ws-ijmpe}
\bibliography{isotope-references-etal}

\end{document}